\documentclass[twoside,reqno]{qts-proc}
\usepackage{epsfig,cite}
\usepackage{amssymb,amsmath}
\usepackage{times}
\begin{document}
\sloppy \raggedbottom
\setcounter{page}{1}
%%%%%%%%%%%%%%%%%%%%%

\newpage
\setcounter{figure}{0}
\setcounter{equation}{0}
\setcounter{footnote}{0}
\setcounter{table}{0}
\setcounter{section}{0}
\def\to{\rightarrow}
\def\gev{\mbox{GeV}}
\def\ev{\mbox{eV}}
\def\mev{\mbox{MeV}}
\def\tev{\mbox{TeV}}
\def\cm{\mbox{cm}}
\def\mpc{\mbox{Mpc}}

\def\AJ{{Astrophys.\ J.\ }}
\def\AJL{{Ap.\ J.\ Lett.\ }}
\def\AJS{{Ap.\ J.\ Supp.\ }}
\def\AM{{Ann.\ Math.\ }}
\def\AP{{Ann.\ Phys.\ }}
\def\APJ{{Ap.\ J.\ }}
\def\APP{{Acta Phys.\ Pol.\ }}
\def\ASJ{{Astron.\ J.\ }}
\def\ASAS{{Astron.\ and Astrophys.\ }}
\def\BAMS{{Bull.\ Am.\ Math.\ Soc.\ }}
\def\CMJ{{Czech.\ Math.\ J.\ }}
\def\CMP{{Commun.\ Math.\ Phys.\ }}
\def\CQG{{Class.\ Quant.\ Grav.\ }}
\def\EURO{{The European Phys. J.}}
\def\FP{{Fortsch.\ Phys.\ }}
\def\GRG{{Gen.\ Rel.\ Grav.\ }}
\def\HPA{{Helv.\ Phys.\ Acta} }
\def\IJMP{{Int.\ J.\ Mod.\ Phys.\ }}
\def\JMM{{J.\ Math.\ Mech.\ }}
\def\JHEP{{JHEP} }
\def\JP{{J.\ Phys.\ }}
\def\JCP{{J.\ Chem.\ Phys.\ }}
\def\LNC{{Lett.\ Nuovo Cimento} }
\def\SNC{{Suppl.\ Nuovo Cimento} }
\def\MPL{{Mod.\ Phys.\ Lett.\ }}
\def\MNRAS{{Mon.\ Not.\ R.\ Ast.\ Soc.\ }}
\def\NAT{{Nature (London)} }
\def\NAST{{New Astronomy} }
\def\NC{{Nuovo Cim.\ }}
\def\NP{{Nucl.\ Phys.\ }}
\def\PL{{Phys.\ Lett.\ }}
\def\PR{{Phys.\ Rev.\ }}
\def\PRD{{Phys.\ Rev.\ D} }
\def\PRL{{Phys.\ Rev.\ Lett.\ }}
\def\PRTS{{Physics Reports} }
\def\PS{{Physica Scripta} }
\def\PTP{{Progr.\ Theor.\ Phys.\ }}
\def\RMPA{{Rev.\ Math.\ Pure Appl.\ }}
\def\RMP{{Rev.\ Mod.\ Phys.\ }}
\def\RNC{{Rivista del Nuovo Cimento} }
\def\SC{{Science} }
\def\SJPN{{Soviet J.\ Part.\ Nucl.\ }}
\def\SP{{Soviet.\ Phys.\ }}
\def\TMF{{Teor.\ Mat.\ Fiz.\ }}
\def\TMP{{Theor.\ Math.\ Phys.\ }}
\def\YF{{Yadernaya Fizika} }
\def\ZETF{{Zh.\ Eksp.\ Teor.\ Fiz.\ }}
\def\ZP{{Z.\ Phys.\ }}
\def\ZMP{{Z.\ Math.\ Phys.\ }}

\def\al{\alpha}
\def\be{\beta}
\def\ga{\gamma}
\def\de{\delta}
\def\ep{\epsilon}
\def\ve{\varepsilon}
\def\ze{\zeta}
\def\et{\eta}
\def\th{\theta}
\def\vt{\vartheta}
\def\io{\iota}
\def\ka{\kappa}
\def\la{\lambda}
\def\vpi{\varpi}
\def\rh{\rho}
\def\vr{\varrho}
\def\si{\sigma}
\def\vs{\varsigma}
\def\ta{\tau}
\def\up{\upsilon}
\def\ph{\phi}
\def\vp{\varphi}
\def\ch{\chi}
\def\ps{\psi}
\def\om{\omega}
\def\Ga{\Gamma}
\def\De{\Delta}
\def\Th{\Theta}
\def\La{\Lambda}
\def\Si{\Sigma}
\def\Up{\Upsilon}
\def\Ph{\Phi}
\def\Ps{\Psi}
\def\Om{\Omega}
\def\cA{{\cal A}}
\def\cB{{\cal B}}
\def\cE{{\cal E}}
\def\cl{{\mathcal L}}
\def\cL{{\mathcal L}}
\def\cO{{\mathcal O}}
\def\cP{{\cal P}}
\def\cR{{\cal R}}
\def\cV{{\cal V}}
\def\mn{{\mu\nu}}

\def\fr#1#2{{{#1} \over {#2}}}
\def\half{{\textstyle{1\over 2}}}
\def\quar{{\textstyle{1\over 4}}}
\def\frac#1#2{{\textstyle{{#1}\over {#2}}}}

\def\vev#1{\langle {#1}\rangle}
\def\bra#1{\langle{#1}|}
\def\ket#1{|{#1}\rangle}
\def\bracket#1#2{\langle{#1}|{#2}\rangle}
\def\expect#1{\langle{#1}\rangle}
\def\norm#1{\left\|{#1}\right\|}
\def\abs#1{\left|{#1}\right|}

\def\lsim{\mathrel{\rlap{\lower4pt\hbox{\hskip1pt$\sim$}}
    \raise1pt\hbox{$<$}}}
\def\gsim{\mathrel{\rlap{\lower4pt\hbox{\hskip1pt$\sim$}}
    \raise1pt\hbox{$>$}}}
\def\sqr#1#2{{\vcenter{\vbox{\hrule height.#2pt
         \hbox{\vrule width.#2pt height#1pt \kern#1pt
         \vrule width.#2pt}
         \hrule height.#2pt}}}}
\def\square{\mathchoice\sqr66\sqr66\sqr{2.1}3\sqr{1.5}3}

\def\prt{\partial}
\def\lrpartial{\raise 1pt\hbox{$\stackrel\leftrightarrow\partial$}}
\def\lrprt{\stackrel{\leftrightarrow}{\partial}}
\def\lrprtnu{\stackrel{\leftrightarrow}{\partial^\nu}}
\def\lrdla{\stackrel{\leftrightarrow}{D^\la}}
\def\lrvec#1{ \stackrel{\leftrightarrow}{#1} }

\def\Re{\hbox{Re}\,}
\def\Im{\hbox{Im}\,}
\def\Arg{\hbox{Arg}\,}

\def\hb{\hbar}
\def\etal{{\it et al.}}

\def\pt#1{\phantom{#1}}
\def\ni{\noindent}
\def\ol#1{\overline{#1}}
\def\s{\scriptstyle}
\def\ss{\scriptscriptstyle}

\def\a{$a_\mu$}
\def\b{$b_\mu$}
\def\d{$d_{\mu\nu}$}
\def\e{$e_\mu$}
\def\f{$f_\mu$}
\def\g{$g_{\la\mu\nu}$}
\def\H{$H_{\mu\nu}$}
\def\aa{a_\mu}
\def\bb{b_\mu}
\def\cc{c_{\mu\nu}}
\def\dd{d_{\mu\nu}}
\def\ee{e_\mu}
\def\ff{f_\mu}
\def\ggg{g_{\la\mu\nu}}
\def\HH{H_{\mu\nu}}

\newcommand{\beq}{\begin{equation}}
\newcommand{\eeq}{\end{equation}}
\newcommand{\bea}{\begin{eqnarray}}
\newcommand{\eea}{\end{eqnarray}}
\newcommand{\rf}[1]{(\ref{#1})}

\def\cosec{\mathop{\rm cosec}\nolimits}
\def\sech{\mathop{\rm sech}\nolimits}
\def\cosech{\mathop{\rm cosech}\nolimits}


% use the thanks command within \title, \author or \address for footnotes:
% \title{Title} or  \title{Title\thanks{...}}

\title{Aspects of spacetime-symmetry violations}

\runningheads{Ralf Lehnert}{Aspects of spacetime-symmetry violations}

\begin{start}

% \author{Name1}{aff.label1}, \coauthor{Name2}{aff.label2},  \coauthor{Name3}{aff.label3}
% \address{Address1}{aff.label1}\address{Address2}{aff.label2}\address{Address3}{aff.label3}

\author{Ralf Lehnert}{}
%\coauthor{Name2}{2},
%\coauthor{Name3}{3}

\address{Department of Physics and Astronomy, Vanderbilt University, Nashville, TN 37235, USA}{}
%\address{Address2}{2}
%\address{Address3}{3}

%you may repeat \coauthor as many times as you need
%names may have more than one aff.label, e.g.,
%\coauthor{Name2}{aff.label2,aff.label3},

\begin{Abstract}
The violation of spacetime symmetries provides a promising
candidate signal for underlying physics, possibly arising at
the Planck scale. This talk gives an overview over various
aspects in the field, including some mechanisms for Lorentz
breakdown, the SME test framework, and phenomenological
signatures for such effects. 
\end{Abstract}
\end{start}

%%%%%%%%%%%%%%%%%%%%%%%%%%%%%%%%%%%%%%%%%%%%%%%%%%%%%%%%%%%%%
% The main text of your paper                               %
%%%%%%%%%%%%%%%%%%%%%%%%%%%%%%%%%%%%%%%%%%%%%%%%%%%%%%%%%%%%%
\section{Introduction}
\label{intro}

Although phenomenologically successful,
the Standard Model of particle physics
leaves unanswered a variety of theoretical questions.
At present,
significant theoretical work
is therefore directed 
toward the search 
for an underlying theory 
that includes a quantum description of gravity.
However,
observational tests of such ideas
face a major obstacle of practical nature:
most quantum-gravity effects 
in virtually all leading candidate models 
are expected to be extremely small
due to Planck-scale suppression.

During the last decade,
minuscule violations of Lorentz and CPT invariance
have been identified as promising
Planck-scale signals \cite{cpt04}.
The basic idea is 
that these symmetries hold exactly in established physics,
are amenable to ultrahigh-precision tests, 
and may be broken in a number of approaches to quantum gravity.
As examples, 
we mention
strings \cite{kps},
spacetime foam \cite{ell98,suv},
nontrivial spacetime topology \cite{klink},
loop quantum gravity \cite{amu},
noncommutative geometry \cite{chklo},
and cosmologically varying scalars \cite{varying}.

The low-energy effects associated with Lorentz and CPT violation
are described by the Standard-Model Extension (SME) \cite{sme}.
The SME is an effective field theory
at the level of the usual Standard Model
and general relativity.
Its flat-spacetime limit 
has provided the basis 
for modern experimental \cite{mat05} and theoretical investigations
of Lorentz and CPT violation
involving 
mesons \cite{hadronexpt,kpo,hadronth,ak},
baryons \cite{ccexpt,spaceexpt,cane},
electrons \cite{eexpt,eexpt2,eexpt3},
photons \cite{photon},
muons \cite{muons},
and the Higgs sector \cite{higgs}.
It is interesting to note
that neutrino-oscillation experiments
offer discovery potential \cite{sme,neutrinos,nulong}.

The present talk
discusses some topics in this field of research.
In Sec.\ \ref{mech},
we briefly review various mechanisms for Lorentz violation
that have been proposed in the literature.
We specifically focus on Lorentz breaking
through cosmologically varying scalars:
this effect highlights
the interplay of translation and rotation/boost invariance.
It is also  
phenomenologically interesting 
because many cosmological models contain novel scalar fields
with time dependencies driven by the expansion of the universe.
An explicit example of such a model
motivated by $N=1$ supergravity
is given in Sec.\ \ref{sugra}.
This talk is summarized
in Sec.\ \ref{sum}.

\section{Some mechanisms for Lorentz violation}
\label{mech}

Lorentz breaking can occur
in a variety of candidate underlying models.
This section
gives a brief overview of a subset of theoretical ideas
along these lines.
We focus on the mechanisms for Lorentz violation
mentioned in the introduction.
Those (and most other) models
are based on a completely Lorentz-invariant Lagrangian;
symmetry breakdown occurs
because the ground-state solution of the respective equations of motion
does not exhibit Lorentz invariance.
This leads to various immediate consequences.
For example,
spacetime remains Lorentzian,
so that different inertial coordinate systems
are still linked by the usual Lorentz transformations.
Moreover,
conventional spinors and tensors still represent physical quantities.
However,
the vacuum contains a structure
that acts like a background field
selecting a preferred direction.
Then,
the outcome of an experiment
can depend on the orientation
and velocity of the laboratory
implying the violation of particle Lorentz symmetry.

{\em Spontaneous Lorentz and CPT violation in string theory.}
From a theoretical perspective,
spontaneous symmetry breaking (SSB)
is a particularly attractive mechanism
for Lorentz violation.
SSB is experimentally well established
in condensed-matter systems,
and in the electroweak model
it is responsible for mass generation.
The basic idea is
that a symmetric zero-field state
is not the lowest energy configuration.
Non-zero vacuum expectation values (VEV) are,
in fact, 
more favorable energetically.
Within the field theory of the open bosonic string,
it has been demonstrated \cite{kps}
that SSB can trigger VEVs of vector and tensor fields,
which would then select preferred spacetime directions.
There is also theoretical evidence
indicating the presence of spontaneous Lorentz violation
in relativistic point-particle field theories
with nonpolynomial interactions \cite{ak05}.

{\em Spacetime foam.}
The basic idea behind this mechanism is 
that Planck-scale fluctuations
could result in a sea of microscopic virtual black holes
and other topologically nontrivial
spacetime configurations in the vacuum.
Besides violations of conventional unitary quantum mechanics,
this could lead to Lorentz-breaking dispersion relations
for particles propagating in such backgrounds.
The emergence of Lorentz violation
is intuitively reasonable
because the thermal black-hole sea has a rest frame,
which selects a preferred (timelike) direction.
In a subset of these approaches,
the dispersion-relation modifications
are interpreted
as resulting from recoil effects on quantum matter
in such black-hole or D-particle backgrounds \cite{ell98}

{\em Nontrivial spacetime topology.}
This approach studies the physics
resulting from the compactification
of one of the three spatial dimensions \cite{klink}.
On observational grounds,
the compactification radius
must be very large.
Note also 
that the local structure of flat Minkowski space is maintained. 
The finite size of the compactified dimension 
leads to periodic boundary conditions, 
which implies a discrete momentum spectrum 
in this direction 
and a Casimir-type vacuum. 
It is then intuitively reasonable 
that this vacuum possesses a preferred direction
along the compactified dimension.

{\em Loop quantum gravity.}
Another idea how Lorentz violation can arise
has been investigated in loop quantum gravity.
To analyze the the classical limit of the theory,
one considers coherent states
peaked around the classical solution for the metric. 
However, 
one can only take into consideration coherent states 
that do not oscillate at transplanckian scales
where Einstein's theory of gravitation is known to be invalid.
This procedure introduces an absolute distance into such classical limits, 
which is incompatible with special relativity.
As a sample consequence,
the Maxwell equations are modified 
leading to a Lorentz-breaking plane-wave dispersion relation \cite{amu}.

{\em Noncommutative field theory.}
A popular approach to underlying physics
is noncommutative field theory. 
The key idea is that the Minkowski coordinates 
$x^{\mu}$ are no longer ordinary real numbers. 
They are promoted to operators on a Hilbert space
satisfying commutation relations 
of the form $[x^{\mu},x^{\nu}]=i\theta^{\mu\nu}$.
Here, 
$\theta^{\mu\nu}$ 
is a spacetime-constant real-valued tensorial parameter.
The presence of the nondynamical $\theta^{\mu\nu}$ 
in this framework 
typically leads, 
for example, 
to vacuum anisotropies
and is therefore associated with Lorentz violation \cite{chklo}. 

{\em Cosmologically varying scalars.}
A varying scalar,
regardless of the mechanism causing the spacetime dependence,
typically implies the violation of translational invariance.
Since translations and Lorentz transformations
are closely intertwined in the Poincar\'e group,
it is unsurprising
that the translation-symmetry breakdown
can also affect Lorentz invariance.

Consider,
for example,
the angular-momentum tensor $J^{\mu\nu}$,
which generates rotations and Lorentz boosts:
\beq
J^{\mu\nu}=\int d^3x \;\big(\theta^{0\mu}x^{\nu}-\theta^{0\nu}x^{\mu}\big).
\label{gen}
\eeq
Note
that this definition 
contains the energy--momentum tensor $\theta^{\mu\nu}$,
which is no longer conserved
when translational symmetry is violated.
Typically,
$J^{\mu\nu}$
will now exhibit a nontrivial dependence on time,
so that the conventional time-independent 
Lorentz-transformation generators can cease to exist.
As a result,
Lorentz and CPT invariance
are no longer guaranteed.

More intuitively,
the violation of Lorentz symmetry 
in the presence of a varying scalar can be seen as follows.
The 4-gradient of the scalar has to be nonzero
in some region of spacetime. 
This gradient 
then selects a preferred direction in such a spacetime region.
Consider, 
for instance, 
a particle
that has interactions with the scalar.
Then, 
its propagation features
may be different
in the directions perpendicular and parallel to the gradient,
and physically inequivalent directions
signal the violation of rotation invariance.
Since rotations are contained in the Lorentz group,
Lorentz symmetry must be broken.

Lorentz violation induced by spacetime-dependent scalars
can also be established at the level of the Lagrangian.
As an example,
consider
a system with varying coupling $\xi(x)$
and two scalar fields $\phi$ and $\Phi$,
such that the Lagrangian $\mathcal{L}$ includes a kinetic-type term
$\xi(x)\,\partial^{\mu}\phi\,\partial_{\mu}\Phi$.
A partial integration of the action of this system
(e.g., with respect to the first partial derivative in the above term)
leaves unaffected the equations of motion.
The resulting equivalent Lagrangian $\mathcal{L}'$ then contains a term
\begin{equation}
\mathcal{L}'\supset -K^{\mu}\phi\,\partial_{\mu}\Phi,
\label{example}
\end{equation}
where $K^{\mu}\equiv\partial^{\mu}\xi$ is an external
nondynamical 4-vector,
which clearly breaks Lorentz invariance.
Note
that for spacetime dependencies of $\xi$ on cosmological scales,
such as the claimed variation of the fine-structure parameter \cite{Webb},
$K^{\mu}$ is constant to an excellent approximation 
locally---say on solar-system scales. 

\section{Example: a supergravity cosmology}
\label{sugra}

In this section, 
we illustrate the above results
within a specific supergravity model 
that generates the variation of two scalars $A$ and $B$
in a cosmological context.
It results in a fine-structure parameter $\alpha$
and an electromagnetic $\theta$ angle
that depend on spacetime. 
Our analysis is performed
within the framework of $N=4$ supergravity
in four spacetime dimensions.
Although this model is unrealistic in detail,
one can gain qualitative insights into candidate fundamental physics
because it is a limit of  $N=1$ supergravity 
in eleven dimensions,
which is contained in M-theory.

When only one graviphoton $F^{\mu\nu}$
is excited,
the bosonic part of the pure $N=4$ supergravity Lagrangian
is determined by \cite{cj}
\bea
\kappa \cl_{\rm sg}
&=&
-\frac 1 2 \sqrt{g} R
+\sqrt{g} ({\prt_\mu A\prt^\mu A + \prt_\mu B\prt^\mu B})/{4B^2}
\nonumber\\
&&
\qquad\!
-\frac 1 4 \ka \sqrt{g} M F_{\mu\nu} F^{\mu\nu}
-\frac 1 4 \ka \sqrt{g} N F_{\mu\nu} \tilde{F}^{\mu\nu}\; .
\label{lag2}
\eea
Here, the $M$ and $N$ are functions of the scalars $A$ and $B$ given by
\beq
M =\fr
{B (A^2 + B^2 + 1)}
{(1+A^2 + B^2)^2 - 4 A^2}\; ,
\quad
N = \fr
{A (A^2 + B^2 - 1)}
{(1+A^2 + B^2)^2 - 4 A^2}\; .
\label{N}
\eeq
The dual field-strength tensor is denoted by 
$\tilde{F}^{\mu\nu}=\varepsilon^{\mu\nu\rho\sigma}F_{\rho\sigma}/2$,
and $g=-\det (g_{\mu\nu})$.
In what follows,
we rescale $F^{\mu\nu}\to F^{\mu\nu}/\sqrt{\kappa}$,
so that the gravitational coupling $\kappa$
disappears in the equations of motion.

The next step is to gauge 
the internal SO(4) symmetry
of the full $N=4$ supergravity Lagrangian.
This supports the interpretation of $F^{\mu\nu}$ 
as the electromagnetic field-strength tensor.
The resulting potential for the scalars $A$ and $B$
is known to be unbounded from below \cite{dfr77}.
However,
we take a phenomenological approach
and assume 
that in a realistic situation
stability must be ensured
by additional fields and interactions.
At leading order, 
we can then model
the potential for the scalars
with mass-type terms:
\beq
\delta\cl=-\half \sqrt{g} (m_A^2 A^2+ m_B^2 B^2)\; .
\label{lagr}
\eeq
We add these terms to our Lagrangian $\cl_{\rm sg}$ 
in Eq.\ \rf{lag2}.

The complete $N=4$ supergravity Lagrangian
also includes fermionic matter \cite{cj}.
In the present cosmological model,
we can represent the fermions
by the energy--momentum tensor $T_{\mu\nu}$ of dust
describing galaxies and other matter:
\beq
T_{\mu\nu} = \rho u_\mu u_\nu \; .
\label{dust}
\eeq
Here,
$\rh$ is the energy density of the matter
and $u^\mu$ is a unit timelike vector
orthogonal to the spatial hypersurfaces,
as usual.

We are now in a position to determine cosmological solutions
of our supergravity model.
We make the usual assumption 
of an isotropic homogeneous
flat $(k=0)$ Friedmann--Robertson--Walker universe
with the conventional line element
\beq
ds^2 = dt^2 - a^2(t)\; (dx^2 + dy^2 + dz^2)\; .
\label{frw}
\eeq
Here, 
$a(t)$ is the scale factor
and $t$ denotes the comoving time.
The assumption of isotropy prohibits our electromagnetic field
from acquiring nonzero expectation values on large scales,
so that we can set $F^{\mu\nu}=0$.
Then,
our cosmological model is governed
by the equations of motion for the scalars $A$ and $B$
and the Einstein equations.
We remark 
that the fermionic matter is uncoupled from the scalars
at tree level,
so that $T_{\mu\nu}$
is approximately conserved by itself.
We then find
$\rho(t) = c_{\rm n}/a^3(t)$,
where $c_{\rm n}$ is an integration constant.

\begin{figure}
\centerline{\epsfxsize=4.1in\epsfbox{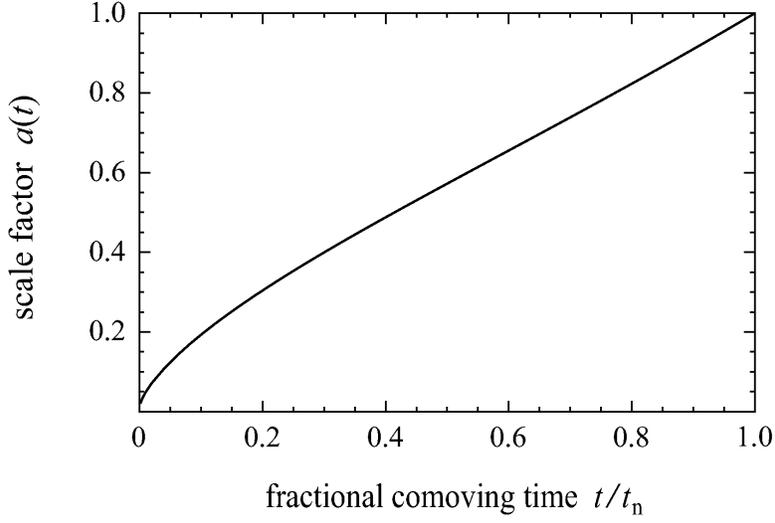}}   
\caption{Scale factor $a(t)$ versus fractional comoving time $t/t_{\rm n}$.
The priors given in Eq.\ \rf{values}
are chosen such that the expansion history of this model
matches closely the one observed for our universe.
\label{sfsol}}
\end{figure}

\begin{figure}
\centerline{\epsfxsize=4.1in\epsfbox{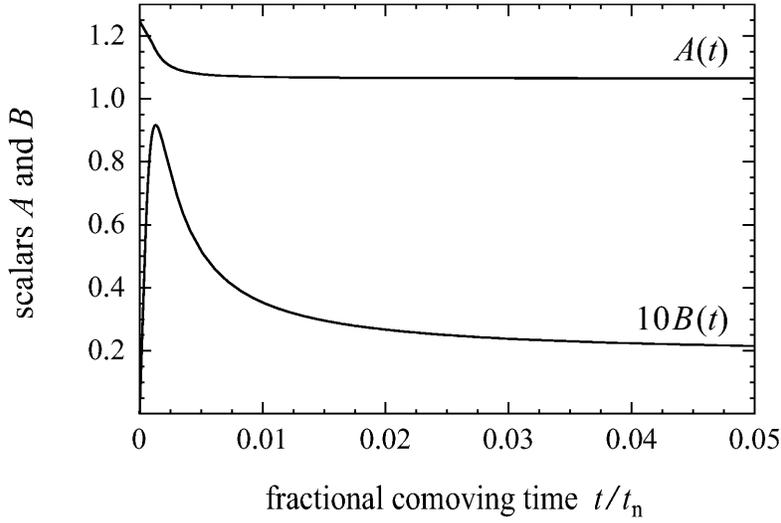}}   
\caption{Time dependence of the scalars $A$ and $B$.
Although at late times the scalars approach constant values,
they do exhibit a nontrivial dependence on the comoving time.
This model is therefore a candidate for exhibiting Lorentz
violation.
\label{ABsol}}
\end{figure}

In special cases,
this cosmological model
admits a variety of analytical solutions \cite{varying}.
In general,
however,
numerical integration is necessary.
A physically interesting scenario
is shown in Figs.\ \ref{sfsol} and \ref{ABsol}.
The input data for this solution are \cite{varying}
\bea
m_A & = & 2.7688 \times 10^{-42}\, {\rm GeV}\; ,\nonumber\\
m_B & = & 3.9765 \times 10^{-41}\, {\rm GeV}\; ,\nonumber\\
c_{\rm n} & = & 2.2790 \times 10^{-84}\, {\rm GeV}^2\; ,\nonumber\\
a(t_{\rm n}) & = & 1 \; ,\nonumber\\
A(t_{\rm n}) & = & 1.0220426 \; ,\nonumber\\
\dot{A}(t_{\rm n}) & = & -8.06401\times 10^{-46}\, {\rm GeV}\; ,\nonumber\\
B(t_{\rm n}) & = & 0.016598 \; ,\nonumber\\
\dot{B}(t_{\rm n}) & = & -2.89477\times 10^{-45}\, {\rm GeV}\; ,
\label{values}
\eea
where the dot denotes differentiation
with respect to the comoving time,
and the subscript n indicates the present value of the quantity.

For purposes of this talk,
the details of this particular solution are less interesting.
It is important to note,
however,
that the scalars $A$ and $B$ 
have acquired a nontrivial dependence on the comoving time $t$:
they vary on cosmological scales.
Thus,
we have established the first requirement for observable Lorentz violation.
Note 
that this feature is common 
to many other, more realistic cosmological models.

Next,
consider
excitations of $F_{\mu\nu}$
in the background cosmological solution
$A_{\rm b}$ and $B_{\rm b}$,
which is depicted in Fig.\ \ref{ABsol}.
Experiments are often confined to small spacetime regions,
so it is appropriate to work in a local inertial frame.
In such a frame, 
the effective Lagrangian $\cl_{\rm cosm}$ for localized $F_{\mu\nu}$ fields
follows from Eq.\ \rf{lag2}
\beq
\cl_{\rm cosm}=
-\frac 1 4 M_{\rm b}F_{\mu\nu} F^{\mu\nu}
-\frac 1 4 N_{\rm b} F_{\mu\nu} \tilde{F}^{\mu\nu}\; .
\label{efflag}
\eeq
Here, 
$M_{\rm b}$ and $N_{\rm b}$
are determined 
by the time-dependent cosmological solutions $A_{\rm b}$ and $B_{\rm b}$. 
Comparison
with the conventional electrodynamics Lagrangian $\cl_{\rm em}$
\beq
\cl_{\rm em} =
-\fr{1}{4 e^2} F_{\mu\nu}F^{\mu\nu}
- \fr{\th}{16\pi^2} F_{\mu\nu} \tilde{F}^{\mu\nu}\; .
\label{em}
\eeq
shows
that $e^2 \equiv 1/M_{\rm b}$ and  $\th \equiv 4\pi^2 N_{\rm b}$.
Since $M_{\rm b}$ and $N_{\rm b}$
depend on the varying-scalar background $A_{\rm b}$ and $B_{\rm b}$,
the electromagnetic couplings $e$ and $\th$
are no longer constant in general.
In light of the Webb data set \cite{Webb},
a time-dependent fine-structure parameter $\alpha$
is intriguing by itself.
However,
here we are interested in the fact
that our cosmologically varying scalar
is coupled to a conventional Standard-Model particle---the photon.
Thus,
the second requirement for observable Lorentz violation is satisfied.

To establish the breakdown of Lorentz symmetry
in our effective electrodynamics 
more clearly,
we can look at the modified Maxwell equations
resulting from Lagrangian \rf{efflag}:
\beq
\fr{1}{e^2}\partial^{\mu}F_{\mu\nu}
-\fr{2}{e^3}(\partial^{\mu}e)F_{\mu\nu}
+\fr{1}{4\pi^2}(\partial^{\mu}\th)\tilde{F}_{\mu\nu}=0\; .
\label{Feom}
\eeq
In our cosmological supergravity model,
the gradients of $e$ and $\th$
appearing in Eq.\ \rf{Feom}
are nonzero,
approximately constant in local inertial frames,
and act like a nondynamical external background.
This vectorial background
selects a preferred direction
in the local inertial frame
breaking Lorentz invariance.

We remark
that the term containing the gradient of $\th$
can be identified 
with a Chern--Simons-type contribution to our modified electrodynamics.
Such a term,
which is included in the minimal SME,
has received substantial attention recently \cite{mcs}.
For instance,
it typically leads to vacuum \v{C}erenkov radiation \cite{cer}.
We also point out
that a Lorentz-violating Chern--Simons-type term for gravity
can be constructed \cite{jp03}.
This term can be generated in a model similar to ours,
which also contains a cosmologically varying scalar \cite{varying}. 

\section{Summary}
\label{sum}

This talk has discussed
various aspects of spacetime-symmetry violations.
The idea is
that various approaches to quantum gravity
can lead to Lorentz-violating ground states,
which are characterized by backgrounds
that select one or more preferred directions.
We have briefly discussed
a few explicit examples leading to such vacua.
One of these examples involves scalars
with a nontrivial spacetime dependence
on cosmological scales.
We have argued
that the involved breakdown of translational invariance
is typically associated Lorentz violation.
This specific mechanism
might be of particular interest
in light of recent cosmological models involving scalar fields
and recent claims of a variation of the fine-structure parameter.

\section*{Acknowledgments}
The author thanks  V.K.\ Dobrev for the invitation
to this stimulating meeting.
Travel support by the NSF is gratefully acknowledged.

\end{document}